\documentclass[conference]{IEEEtran}
\usepackage{amsfonts, amssymb, amsmath, stmaryrd, etoolbox, bbm, amsthm}
\usepackage{bm}
\usepackage{algorithmic}
\usepackage{algorithm}
\usepackage{array}

\usepackage{subcaption}
\usepackage{textcomp}
\usepackage{stfloats}
\usepackage[normalem]{ulem}
\usepackage{url}
\usepackage{verbatim}
\usepackage[usenames,dvipsnames]{xcolor}
\usepackage{graphicx}
\usepackage{cite}
\usepackage{acronym}
\usepackage{hyperref}
\acrodef{PST}[PST]{Phase Shift Transformer}
\acrodef{HVDC}[HVDC]{High Voltage Direct Current}
\acrodef{MDP}[MDP]{Markov Decision-Process}
\acrodef{L2RPN}[L2RPN]{Learn to run a power network}
\acrodef{GPU}[GPU]{Graphics Processing Unit}
\acrodef{GPUs}[GPUs]{Graphics Processing Units}
\acrodef{CPU}[CPU]{Central Processing Unit}
\acrodef{SIMD}[SIMD]{Single Instruction, Multiple Data}
\acrodef{RL}[RL]{Reinforcement Learning}
\acrodef{OR}[OR]{Operations Research}
\acrodef{QD}[QD]{Quality Diversity}
\acrodef{PTDF}[PTDF]{\emph{Power Transfer Distribution Factors}}
\acrodef{LODF}[LODF]{\emph{Line Outage Distribution Factors}}
\acrodef{MODF}[MODF]{\emph{Multiple Outage Distribution Factors}}
\acrodef{BSDF}[BSDF]{\emph{Bus Split Distribution Factors}}
\acrodef{PSDF}[PSDF]{\emph{Phase Shifter Distribution Factors}}
\acrodef{DCDF}[DCDF]{\emph{Direct Current Distribution Factors}}
\acrodef{OOM}[OOM]{Orders of Magnitude}

\theoremstyle{definition}
\newtheorem{definition}{Definition}

\newcommand{\R}{\mathrel{\ \mathbb{R}}}

\newcommand{\vect}[1]{{#1}}

\begin{document}

\title{Accelerated DC loadflow solver for topology optimization}

\author{\IEEEauthorblockN{Nico Westerbeck}
\IEEEauthorblockA{Elia Group \\ Instadeep}
\and
\IEEEauthorblockN{Joost van Dijk}
\IEEEauthorblockA{TenneT Netherlands}
\and
\IEEEauthorblockN{Jan Viebahn}
\IEEEauthorblockA{TenneT Netherlands}
\and
\IEEEauthorblockN{Christian Merz}
\IEEEauthorblockA{Elia Group}
\and
\IEEEauthorblockN{Dirk Witthaut}
\IEEEauthorblockA{Forschungszentrum Jülich}
}

\markboth{Journal of \LaTeX\ Class Files,~Vol.~14, No.~8, August~2021}%
{Shell \MakeLowercase{\textit{et al.}}: A Sample Article Using IEEEtran.cls for IEEE Journals}

\IEEEpubid{0000--0000/00\$00.00~\copyright~2021 IEEE}

\maketitle

\begin{abstract} We present a massively parallel solver that accelerates DC loadflow computations for power grid topology optimization tasks. Our approach leverages low-rank updates of the Power Transfer Distribution Factors (PTDFs) to represent substation splits, line outages, and reconfigurations without ever refactorizing the system. Furthermore, we implement the core routines on Graphics Processing Units (GPUs), thereby exploiting their high-throughput architecture for linear algebra. A two-level decomposition separates changes in branch topology from changes in nodal injections, enabling additional speed-ups by an in-the-loop brute force search over injection variations at minimal additional cost. We demonstrate billion-loadflow-per-second performance on power grids of varying sizes in workload settings which are typical for gradient-free topology optimization such as Reinforcement Learning or Quality Diversity methods. While adopting the DC approximation sacrifices some accuracy and prohibits the computation of voltage magnitudes, we show that this sacrifice unlocks new scales of computational feasibility, offering a powerful tool for large-scale grid planning and operational topology optimization.
\end{abstract}


\section{Introduction}

\IEEEPARstart{P}{ower} system simulation and optimization face a trade-off between accuracy and speed. Many applications, including large-scale grid planning and optimization, require approximations to be computationally feasible~\cite{horsch2018linear}. The DC approximation is widely used in steady-state power system analysis, as it reduces the power flow equations to a system of linear equations~\cite{wood2013power}. Changes in the grid topology, such as line outages of bus reassignments, can be described by low-rank updates of the system of linear equations~\cite{modf,BSDF}. If the workload holds large batches of similar problems, such updates are promising candidates to speed up the loadflow solving process. Furthermore, modern \ac{GPUs} can drastically speed up computations in linear algebra and thus may further alleviate the computational bottleneck. In this article, we present a novel solver architecture designed to solve large batches of similar loadflow problems without ever re-inverting the system of linear equations. We implement our solver on CPU and on modern GPU hardware with slightly different approaches. We tailor this solver for applications in topology optimization, where many small changes to the grid in the form of bus splits, reassignments, or branch disconnections must be evaluated. Benchmarked on multiple test grids, we demonstrate a performance on the order of billions of loadflows per second.

The developed solution harnesses the potential of modern \ac{GPU} hardware. Originally designed to accelerate graphics rendering for computer games, \acs{GPU} have been widely used for various accelerated computing tasks since the introduction of CUDA~\cite{cuda}. Their vector processors allow for massively parallel execution of code through the \ac{SIMD} paradigm. In this paradigm, every vector processor executes the same instruction in a clock cycle under varying data inputs. Matrix-multiplication tasks benefit from extra speedups through systolic array processors such as NVIDIAs tensor cores~\cite{tensorcores}.

The prime use-case for the developed loadflow solver is topology optimization as an alternative to costly congestion management methods. Congestion is becoming increasingly important in modern grids with a high penetration of renewable power sources (see, e.g.,~\cite{titz2024identifying}). Grid operators typically apply redispatch, shifting generation from one side of a grid bottleneck to the other. Congestion can be substantially mitigated by low-cost switching actions that adjust the topology to reroute flows~\cite{subramanian2021exploring}. In our setting, we consider three types of topology actions:
\begin{enumerate}
\item Line disconnections
\item Opening of a busbar coupler
\item Modifying the busbar assignment, both for branches and injections
\end{enumerate}
Furthermore, the topology can be optimized by changing \ac{PST} taps or \ac{HVDC} setpoints. These measures are beyond the scope of this work as there are existing approaches to optimize these~\cite{CAPITANESCU20111731}.
Other remedial actions, such as closing of busbar couplers and reconnection of lines, are deemed of less importance, as the operational planning foresees open couplers or lines typically only in case of maintenance. Hence, a reconnection may be costly of even important.

Fundamentally, there are two ways to approach topology optimization. A gradient based approach within the field of \ac{OR} methods such as Linear Programming to directly optimize for switching actions \cite{Preuschoff:996209}, or gradient free optimization like \ac{RL} \cite{grid2op, dorfer2022powergridcongestionmanagement, pmlr-v133-marot21a, 9494879} or \ac{QD} which can query an oracle for the quality of a proposed topologies. For the scope of this work, we focus on the latter, which requires a large amount of loadflow computations to be successful. 

Our article is organized as follows: In Section ~\ref{sec:background} we present the mathematical background to the paper. Section ~\ref{sec:approach} presents our implementation on CPU and on GPU. Section~\ref{sec:design} discusses some trade-offs in the software architecture and the respective design choices. Section~\ref{sec:results} benchmarks the solution in terms of performance on multiple test grids. Section~\ref{sec:Discussion} discusses further work.

\section{Background: DC Loadflow Theory}
\label{sec:background}

This section introduces the fundamentals of the DC approximation with a focus on topology changes. Any topology change can be described by a low-rank change to the grid Laplacian or nodal susceptance matrix, respectively. If this Laplacian has been inverted once for a reference topology, other topologies can be evaluated efficiently by updating the reference solution. These low-rank updates boil down to mainly matrix multiplications, making it an excellent candidate for GPU acceleration. 

We use bracket index notation with $x_{[i]}$ as the vector $x$ at index $i$ and $X_{[i, j]}$ as the matrix $X$ in the $i$th row, $j$th column in this work.

The DC loadflow is a linearization of the AC loadflow. For a DC loadflow, the network is modeled as a weighted directed graph $G = (V, E)$, where $V$ denote the set of vertices or nodes and $E$ denote the set of edges. For convenience we simply write $V = \{0, \dots, |V|-1\}$, $E = \{0, \dots, |E|-1\}$. The direction of the edges is defined using a from-vector $\vect{f} \in \R^{|E|}$ and a to-vector $\vect{t} \in \R^{|E|}$. We write $f_k$ or $t_k$ for indexing the $k$th entry in $\vect{f}$ or $\vect{t}$. The susceptances are weights given by a weight vector $\vect{b} \in \R^{|E|}$.

\begin{definition}
% Every network $G$ has a \emph{connectivity matrix} $\vect{C} \in \R^{|E| \times |V|}$ given by
% \begin{align*}
%     \vect{C}_{\ell n} =   \begin{cases}
%         1  & \text{if }  \vect{f}_{\ell} = n, \\
%         -1 & \text{if }  \vect{t}_{\ell} = n, \\
%         0  & \text{else}. 
%   \end{cases} && \text{for } \ell \in E, n \in V.
% \end{align*}
The \emph{weighted connectivity matrix} $\vect{C}_w \in \R^{|E| \times |V|}$ is given by
\begin{align*}
    \vect{C}_{w[\ell, n]} =   \begin{cases}
        \vect{b}_{[\ell]}  & \text{if }  \vect{f}_{[\ell]} = n, \\
        -\vect{b}_{[\ell]} & \text{if }  \vect{t}_{[\ell]} = n, \\
        0  & \text{else}. 
  \end{cases}  && \text{for } \ell \in E, n \in V.
\end{align*}
The \emph{nodal susceptance matrix} $\vect{B} \in \R^{|V| \times |V|}$ is a graph Laplacian given by
\begin{align*}
    \vect{B}_{[n, m]} =   \begin{cases}
        -\vect{b}_{[(n, m)]}  & \text{if } (n, m) \in E_n \text{ or } (m, n) \in E_n, \\
        \sum_{\ell \in E_n}\vect{b}_{[\ell]} & \text{if }  n=m, \\
        0  & \text{else}. 
  \end{cases}  
\end{align*}
for $n,m \in V$ and where $E_n$ denotes the set of edges incident on the node $n$
\end{definition}

A node need to be designated as the \emph{slack node}. Let the prime symbol $(\cdot)'$ denote the exclusion of the slack node from the rows and/or columns. For instance in $\vect{B}'$ both the slack row and column are removed, and for $\vect{C}_w'$ the slack column is removed.

\begin{definition}
\label{eq_ptdf}
The \acf{PTDF} \cite{ptdf_psdf_dcdf, ronellenfitsch2016dual, zocca2021spectral, landgren1972transmission} is the solution matrix $\vect{PTDF} \in \R^{|E| \times |V|}$ to the following linear system.
\begin{align}
    \vect{PTDF}' \cdot \vect{B}' = \vect{C}_w' \ \label{eq:ptdf_def}. 
\end{align}
The column of the slack node is defined to be zero.
\end{definition}

In practice, not all rows of the \ac{PTDF} are needed for the calculation. In general, only the edges that are to be monitored, outaged or used in switching/busbar reconfiguration operations are necessary for the calculations described in the paper. Hence, a speedup can be obtained by restricting Eq. \eqref{eq:ptdf_def} to only the necessary edges.

Given the \ac{PTDF} matrix, the N-0 loadflow results can be computed through
\begin{align}
\label{eq_ptdf_times_nodal_inj}
    \vect{p_{n0}} = \vect{PTDF} \cdot \vect{P_n}
\end{align}
where $\vect{P_n} \in \R^{|V|}$ is the nodal active power. For a node $m \in V$, if $I_m$ is the set of injections (e.g. loads and generators) connected to $m$ and $\vect{P_i}$ the setpoint of that injection, then
$\vect{P_{n[m]}} = \sum_{i \in I_m}\vect{P_i}$.

Instead of computing the full loadflows, it is also possible to update a previously computed loadflow $\vect{p_{n0, old}}$ to include a change in injection $\Delta \vect{P_n}$:
\begin{align}
    \label{eq_ptdf_delta}
    \vect{p_{n0}} = \vect{p_{n0, old}} + \vect{PTDF} \cdot \Delta \vect{P_n}
\end{align}
When $\Delta \vect{P_n}$ is sparse, this is more efficient than recomputing the flows through Eq. \eqref{eq_ptdf_times_nodal_inj}. 

In practice, one can exploit these facts to reduce the effective PTDF matrix size. After precomputation, one can eliminate all nodes that always remain static. First, note that only a small selection of nodal power on the nodes will be shifted around in busbar reconfigurations/needed in outage calculations. If $T \subseteq V$ denotes the nodes which are not changed by this, and $\vect{p_T} \in \R^{|E|}$ is defined by
\begin{align}
    \vect{p_{T[\ell]}} = \sum_{m \in T}\vect{PTDF}_{[\ell,m]} \cdot \vect{P_{n[m]}},
\end{align}
then if the nodes of $T$ are removed from the \ac{PTDF} and $p_T$ is added as a new column, with a nodal power of 1 for that column, then the calculations still work.

\begin{definition}
\label{eq_lodf}
The \acf{LODF} \cite{ronellenfitsch2017dual, strake2019non, landgren1972transmission} for the outage of an edge $k \in E$ is given by
\begin{align*}
    \vect{LODF}_{[\ell, k]} = \begin{cases}
        \frac{\vect{PTDF}_{[\ell, \vect{f}_k]} - \vect{PTDF}_{[\ell, \vect{t}_k]}}{1 - (\vect{PTDF}_{[k, \vect{f}_k]} - \vect{PTDF}_{[k, \vect{t}_k]})} & \text{if } \ell \neq k\\
        -1 & \text{if }\ell = k
    \end{cases} \ ,
\end{align*}
where $\ell \in E$ is an edge.
\end{definition}

The post-outage flows $p_{n1}$ after outaging branch $k$ can be obtained through
    \begin{align}
        \label{eq_lodf_delta}
        \vect{p_{n1[k]}} = \vect{p_{n0}} + \vect{LODF}_{[:,k]} \cdot {\vect{p_{n0[k]}}}
    \end{align}

where the subscript $[:, k]$ denotes the $k$th column of the matrix. The matrix that stacks all ${p_{n1[k]}}$ is referred to as N-1 matrix.

Furthermore, the PTDF matrix can be updated to reflect the impact of the outaged branch with
\begin{align}
    \label{eq_lodf_ptdf}
    \vect{PTDF}' = \vect{PTDF} + \vect{LODF}_{[:, k]} \cdot \vect{PTDF}_{[k,:]}
\end{align}
This way, multiple outages can be represented through sequential application of the LODF. However, in \cite{modf, modf2, zocca2021spectral, kaiser2020collective} an alternative formula for multiple outages was developed:

\begin{definition}
\label{eq_modf}
The \acf{MODF} for the outage of edges $\mathcal{O} = \{k_0, \dots, k_N\} \subseteq E$ is the matrix $\vect{MODF} \in \R^{|E| \times (N + 1)}$ given by solving the linear system
\begin{align*}
    &\vect{MODF} \cdot (I - (\vect{PTDF}_{[\mathcal{O}, \vect{f}\mathcal{O}]} - \vect{PTDF}_{[\mathcal{O}, \vect{t}\mathcal{O}]}))
    \\ &= \vect{PTDF}_{[:, \vect{f}\mathcal{O}]} - \vect{PTDF}_{[:, \vect{t}\mathcal{O}]} \ .
\end{align*}
where $\vect{PTDF}_{[A, B]} \in \R^{|A| \times |B|}$, $A \subseteq E$, $B \subseteq V$, is defined by $(\vect{PTDF}_{[A, B]})_{[\ell, n]} = \vect{PTDF}_{[\ell, n]}$, and the sets $\vect{f}\mathcal{O}$, $\vect{t}\mathcal{O}$ by $\vect{f}\mathcal{O} = \{\vect{f}_{\ell}|\ \ell \in \mathcal{O}\}$, and $\vect{t}\mathcal{O} = \{\vect{t}_{\ell}|\ \ell \in \mathcal{O}\}$. That is, we keep only the columns of the PTDF matrix corresponding to the from- and to-nodes of the outaged edges.

After this calculation, we must ensure that the flow on the outages disappears, which can be done by setting
\begin{align*}
    \vect{MODF}_{[k_i, j]} = \begin{cases}
        -1 & \text{if }i = j\\
        0  & \text{if }i \neq j
    \end{cases} 
\end{align*}
for $i, j \in \{0, \dots, N\}$. The \ac{MODF} updates of the \ac{PTDF} and flows works similarly to the \ac{LODF}.

\end{definition}

For the bus split we shall use the \ac{BSDF} \cite{BSDF, vandijk2024unifiedalgebraicdeviationdistribution}. Say that $A \in V$ is the node we want to split, and that $B = |V|+1$ is the new node split off $A$ after the split. Suppose that $S \subseteq E$ is the subset of edges that (are connected and) remain connected to $A$ after the split. 
\begin{definition}
    The \emph{shifted injection vector} $\vect{P}_{bsdf, shift} \in \R^{|V| + 1}$ is defined by
    \[
        \vect{P}_{bsdf, shift[n]} = \begin{cases}
            -1 & \text{if } n=B \\
             \frac{\vect{b}_{\ell}}{\sum_{e \in S}\vect{b}_e}  & \text{if } \ell \in S \text{ s.t. } \vect{f}_{\ell}=n \text{ or }\vect{t}_{\ell}=n\\
            0 & \text{else}
        \end{cases}
    \]
    The \emph{correction vector} $\vect{P}_{bsdf, corr} \in \R^{|E|}$ is defined by
    \[
        \vect{P}_{bsdf, corr[\ell]} = \begin{cases}
             \frac{\vect{b}_{\ell}}{\sum_{e \in S}\vect{b}_e}  & \text{if } \ell \in E \text{ s.t. } \vect{f}_{\ell}=A,  \vect{t}_{\ell}=n \\
             \frac{-\vect{b}_{\ell}}{\sum_{e \in S}\vect{b}_e}  & \text{if } \ell \in E \text{ s.t. } \vect{f}_{\ell}=n,  \vect{t}_{\ell}=A\\
            0 & \text{else}
        \end{cases}
    \]
\end{definition}

\begin{definition}
\label{eq_bsdf}
The \acf{BSDF} is a vector $\vect{BSDF} \in \R^{|E|}$ given by
\begin{align*}
    \vect{BSDF} = %\begin{cases}
        \frac{\vect{PTDF} \cdot \vect{P}_{bsdf, shift} + \vect{P}_{bsdf, corr}}{1 - \vect{PTDF}_{bbc} \cdot \vect{P}_{bsdf, shift}} \ .
    % \end{cases}
\end{align*}
    % \textcolor{red}{Fill in BSDF formula.}
Here, $\vect{PTDF}_{bbc}$ is the calculated formal \ac{PTDF} over the busbar coupler. We emphasize that the busbar coupler is not an edge in $E$. See \cite{BSDF} for details.

For repeated application of the \ac{BSDF} for multiple splits, the \ac{PTDF} can be updated according to
\begin{align}
    \label{eq_bsdf_application}
\vect{PTDF}' = \vect{PTDF} + \vect{BSDF} \cdot \vect{PTDF}_{bbc}
\end{align}
and the formula can be applied repeatedly.
\end{definition}
\acp{PST} and \ac{HVDC} lines can be modelled via \acf{PSDF} and \acf{DCDF} \cite{ptdf_psdf_dcdf}.

%%%%%%%%%%%%%%%%%
\section{Solver Architecture}\label{sec:approach}

We devise a loadflow solver for massively parallel evaluation of topologies under the DC approximation. First, we define the task of the loadflow solver and then present a CPU and a GPU implementation approach, fulfilling that task.

\subsection{Task Definition}\label{TaskDefinition}

A gradient free topology optimizer must evaluate a very large number of candidate topologies for different time intervals and contingency scenarios. Hence, a very fast loadflow solver is the key component of any such optimizer. The task of this solver is to compute real power flows on all monitored branches for every candidate topology. More specifically, we require N-0 and N-1 flows $p_{n0} \in \R^{t\times m}$ and $p_{n1} \in \R^{t \times o \times m}$ with $t$ denoting the time dimension, $o$ the outage dimension and $m$ the monitored branch dimension. To enforce equipment constraints, further quantities of interest could be part of the loadflow results, like the cross-coupler flows during the split and voltage angle differences between open couplers. For the scope of this work, we assume only N-0 and N-1 results to be of interest, while other extensions can be added relatively easily. 

We assume a defined metrics aggregation function $agg_m(p_{n0}, p_{n1}) \rightarrow \R$ which aggregates the loadflow results into a scalar optimization objective. Such an objective could be a combination of constraint penalties. Note that in linear programming, the optimization objective must be differentiable, while in our setting we also allow for non-differential elements such as the number of branches overloaded or the number of switched breakers. For the benchmarks in this work, we minimize maximum relative branch load. 
Furthermore, we specify a user-defined information aggregation function $agg_i(p_{n0}, p_{n1})$ which extracts arbitrary relevant information for presentation to the operators or other parts of the system. For this work, we assume a function that returns sparse N-0 and N-1 results containing only the worst branch results as relevant information. 
Analog to the grid optimization framework grid2op \cite{grid2op} we utilize topology vectors to store a topology. We limit ourselves to two-way splits, hence the topology of a splittable busbar can be expressed as a vector of booleans where every boolean indicates if an element is connected to busbar A (False) or busbar B (True). To facilitate solving, we split this vector into two components, a branch component $t_b \in \{0, 1\}^{|E_c|}$ where $|E_c|$ is the number of branches on either side of the coupler, and an injection component $t_i \in \{0, 1\}^{|I_c|}$ where $|I_c|$ is the number of injections on either side of the coupler.  As injection combinations are easy to re-compute, we devise the option to evaluate a set $T_i$ of injection topologies for every branch topology. A task of the solver is to evaluate $agg_m$ for $p_{n0}$ and $p_{n1}$ obtained through every injection topology in the set and return the best one ($t_i^*$). This integrates an injection bruteforce optimization directly into the solver:
\begin{equation}
    t_i^* = \text{argmin}_{t_i \in T_i} agg_m(t_i)
\end{equation}

Finally, we also add remedial branch disconnections as $t_d \in [0, |E|]^{x}$ as an index vector of branches to be disconnected along with a topology. Note the difference between branch disconnections as remedial actions and branch disconnections as part of the N-1 computation. Branch disconnections as remedial actions need to yield a  N-1 safe grid and are hence treated separately.
We can then define the loadflow solving process as a function:
\begin{equation}
    \label{eq:solver_end_to_end}
    \begin{pmatrix}
        t_b  \\
        t_d  \\
        T_i 
    \end{pmatrix}
    \rightarrow
    \begin{pmatrix}
        agg_m(t_b, t_d, t_i^*) \\
        agg_i(t_b, t_d, t_i^*) \\
        t_i^* 
    \end{pmatrix}
\end{equation}
where $agg(t_b, t_d, t_i^* )$ denotes the evaluation of $agg$ for the given branch topology and disconnections and the found best injection candidate.

Note that we assume a computation in batches. Hence, not only a single $(t_b, t_d, T_i)$ input but a large amount of them is to be computed. While in theory, the search space is large enough to justify practically infinite size of batches, memory limitations will practically limit the maximum batch size and a repeated application of the solver might be necessary.

The task to come up with a selection of candidate topologies to be evaluated is the responsibility of an optimizer and thus beyond the scope for this work. However, the optimizer could employ mechanisms to pre-filter topologies by eliminating symmetries or assignments that disconnect the grid under N-1.

The following sections will devise two approaches how to implement the function defined in Eq. \eqref{eq:solver_end_to_end}, either on CPU or on GPU.

\subsection{CPU Architecture}
\label{sec:cpuapproach}
We devise a CPU baseline for our approach, utilizing the BSDF formulation to speed up computation. Note that for a combination of multiple split busbars, the BSDF formulation from Def.~\ref{eq_bsdf} is used to iteratively implement all splits of a topology. These iterative updates of the PTDF matrix  yield intermediate matrices that can be used to compute other topologies with the same intermediate matrices. Fundamental to the idea behind the CPU approach is a tree construction to avoid duplicate BSDF applications. If a great amount of branch topologies are to be computed, it is likely that some topologies share common splits. Arranging the bus splits in a tree will allow to compute common splits only once at the root of the tree and all ancestors to branch off the same PTDF matrix.

\begin{figure}
\begin{subfigure}{0.45\textwidth}
    \includegraphics[width=0.95\linewidth]{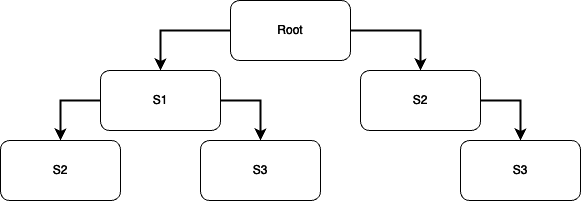}
    \caption{An example tree alignment of the computations, amounting to five BSDF applications. Memory requirement is proportional to the depth of the tree if no parallelization is used.}
    \label{fig:dc_tree}
\end{subfigure}
\begin{subfigure}{0.45\textwidth}
    \includegraphics[width=0.95\linewidth]{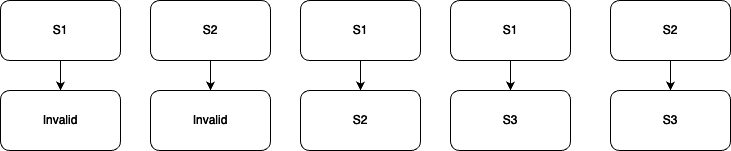}
    \caption{A naive parallelization without the tree idea, amounting to ten BSDF applications but computable in batches of five with memory requirement proportional to the batch size. In this SIMD parallelizable approach, two invalid blocks are inserted, marking unutilized vector processors.}
    \label{fig:gpu_sequence}
\end{subfigure}
\caption{Two different layouts of computation for the set of branch topologies \texttt{S1, S2, S1+S2, S1+S3, S2+S3}. The tree-formulation needs fewer BSDF applications, while the sequence computation exhibits more parallelization potential}
\end{figure}

For example, suppose we have three splittable substations \texttt{S1, S2, S3} and disregard busbar assignments, i.e. substations are always split to the same configuration. Suppose we want to open couplers in the following configurations: \texttt{S1, S2, S1+S2, S1+S3, S2+S3}. Naively, these are eight applications of the BSDF formula. By structuring these computations in a tree as in Fig.~\ref{fig:dc_tree}, the amount of computations can be reduced to five in this example. Parallelizing them in SIMD even virtually increases the number of computations to ten as two vector processors will be unutilized in the second loop. Traversing the tree in a depth-first manner minimizes the amount of memory required, bounding the number of intermediate PTDF matrices to the depth of the tree. Note that if different branch busbar assignments are computed, these are different splits even if they happen on the same substation and would require different tree nodes.

Once the PTDF matrix updated with all splits is computed, the bruteforce search on $T_i$ starts (see Sec.~\ref{TaskDefinition}). The nodal injection vector $P_n$ is updated for every $t_i \in T_i$, giving $p_{n0}$ with Eq.~\ref{eq_ptdf_times_nodal_inj}. As only a handful of injections are changed, Eq.~\ref{eq_ptdf_delta} can be utilized to reduce the size of the matrix multiplication during this iteration process. After computing $p_{n0}$, $p_{n1}$ is obtained using the LODF formulation from Def.~\ref{lodf}. Finally, the best injection topology $t_i^*$ is selected according to $agg_m$ and relevant information is extracted according to $agg_i$.

We implement this approach in python and numpy to establish a CPU baseline performance, as there is currently no CPU implementation of a BSDF based solver known to us. The code can be trivially parallelized over topology inputs, which we do using ray \cite{ray}, splitting the compute set into equal-sized parts for every worker. We omit an implementation of disconnection and multi-outage processing in the CPU baseline.

\subsection{GPU Architecture}\label{sec:gpuapproach}
% \begin{figure}
% \includegraphics[width=0.48\textwidth]{branch_computations.png}
% \caption{Branch computation module}

% \end{figure}

On GPU, the tree approach could be implemented in principle, however there are some difficulties with it. To fully utilize the SIMD capabilities of GPUs, it is desired to perform BSDF applications in parallel. However, the size of the PTDF matrix of a production TSO grid in the hundreds of megabytes memory-bounds the computation. Only a handful of PTDF matrix copies can be stored, limiting the parallelism possible on a tree approach. Furthermore, we envision using the solver in an agent-environment loop as common in \ac{RL} or \ac{QD}. These agent-environment interactions tend to output many distinct looking topologies with few common splits as these algorithms are specifically designed to efficiently explore the search space. Hence, we hypothesize that the performance saving by avoiding recomputations does not outweigh the drawbacks of lower SIMD parallelism and higher complexity. 

Instead, we decide to implement a naive parallelism scheme without the tree formulation on GPU. We parallelize the BSDF computations over all topologies in the input set as shown in Fig.~\ref{fig:gpu_sequence}, evaluating a fixed batch size in SIMD parallel and iterating over topology depth. This bounds the number of PTDF matrices in memory to the batch size, but requires dummy computations to be inserted in the case of varying topology depth. In the example presented, this increases the amount of effective BSDF applications from eight to ten, as the invalid blocks will not free up compute time for other work.

We describe the GPU computation in two components: A branch-computation module and an injection-computation module. The branch module is responsible for all computations that need to be done only once per branch topology. The injection module takes the information from the branch module and for every injection candidate, computes the N-0 and N-1 matrices and invokes $agg_m$ to select $t_i^*$. For $t_i^*$ the output $agg_i$ is stored.

The branch module consists of the following steps, running all computations in SIMD parallel:
\begin{enumerate}
    \item Select a branch topology, disconnection and injection topology set $(t_b, t_d, T_i)$ from the set of all topologies that shall be computed.
    \item \label{bsdf} Update the PTDF matrix for $t_b$ using the BSDF formulation from Def.~\ref{eq_bsdf}, storing the modified PTDF.
    \item \label{disconnection} Update the modified PTDF matrix for static disconnections in $t_d$ using the MODF formulation from Def.~\ref{eq_modf}.
    \item \label{lodf} Compute the LODF matrix for N-1 outages according to Def.~\ref{eq_lodf}.
    \item \label{modf} Compute the MODF matrix for multi-outages that are part of the N-1 computation according to Def.~\ref{eq_modf} for each topology.
    \item \label{static_flows} Compute the static part of the flows that won't change due to injection combinations, as discussed in Def.~\ref{eq_ptdf}.
\end{enumerate}
The injection module consists of these steps:
\begin{enumerate}
    \addtocounter{enumi}{6}
    \item Pre-allocate a buffer for one $agg_i$ output per branch topology.
    \item \label{nodal_injection} Choose a batch of $t_i \in T_i$ that hasn't been computed, and in SIMD parallelism:
    \begin{enumerate}
        \item \label{n0} Compute N-0 flows according to Def.~\ref{eq_ptdf}, using the updated \ac{PTDF} which was computed in the branch module. For this, only the indices in $P_n$ that have changed need to be recomputed, the rest was computed in step~\ref{static_flows}.
        \item \label{n1_lodf} Compute N-1 flows for single-branch outages by applying the LODF matrix from step~\ref{lodf} to the N-0 flows.
        \item \label{n1_modf} Compute N-1 flows for multi-outages by applying the MODF matrix from step~\ref{modf} to the N-0 flows.
        \item \label{n1_inj} Compute N-1 flows for generator failures by utilizing Eq. \eqref{eq_ptdf_delta}. 
        \item \label{metric} Compute a scalar metric to rate injection candidates according to $agg_m$.
        \item \label{aggregate} Aggregate information according to $agg_i$, if $agg_m$ is better than the current best, overwrite the pre-allocated memory. Otherwise, discard.
    \end{enumerate}
\end{enumerate}

This configuration of the injection module will be referred to as \textit{output-first} mode. There is a variation of this procedure which we refer to as \textit{metric-first} mode:
\begin{enumerate}
    \addtocounter{enumi}{6}
    \item Do not create a memory buffer for $agg_i$.
    \item Choose a batch of $t_i \in T_i$ that hasn't been computed, and in SIMD parallelism:
    \begin{enumerate}
        \item[a-e)] Similar to output-first-mode
        \addtocounter{enumii}{5}
        \item Do not aggregate according to $agg_i$.
    \end{enumerate}
    \item Select $t_i^*$.
    \item Recompute $p_{n0}$ and $p_{n1}$ for $t_i^*$. Aggregate according to $agg_i$.
\end{enumerate}

For the special case of $|T_i| = 1$, the code can be simplified, avoiding the bruteforce loop. We refer to that as \textit{symmetric} mode. The difference of these modes is discussed in Sec.~\ref{sec:metric_first}.

\section{Design decisions}
\label{sec:design}
This section introduces the rationale behind multiple architecture decisions set out in  Sec.~\ref{sec:approach}, justifying them either qualitatively or quantitatively.

\subsection{Injection-branch differentiation and injection bruteforcing}
\label{sec:bruteforce}
We differentiate between computations that affect only the susceptances or connectivities in the network and hence have the potential of modifying the \ac{PTDF}-matrix \emph{(branch topologies)} and computations that affect only the nodal injections \emph{(injection topologies)} and are irrespective of the \ac{PTDF} matrix. \Ac{PTDF}-changing actions are reconfiguration of the transformers and lines in a substation and static outages. Reconfiguration of loads or generators in a substation only affects the nodal injections. The effort to compute a PTDF change is higher than a nodal injection change, as outer products are required to update the PTDF with BSDF, LODF or MODF. Especially with larger PTDF matrices, as found in production systems, combined with the $O(n^2)$ complexity, these operations tend to take a significant portion of runtime. An injection reconfiguration under a precomputed PTDF is only a vector-scalar operation to get the N-0 flows (Eq.~\ref{eq_ptdf_delta}), and a single matrix-vector operation to get the N-1 matrix (Eq.~\ref{eq_lodf_delta}). This difference in runtime is worth exploiting by separating injections from branches and performing an in-the-loop search over injection combinations.

\begin{figure}
    \includegraphics[width=0.95\linewidth]{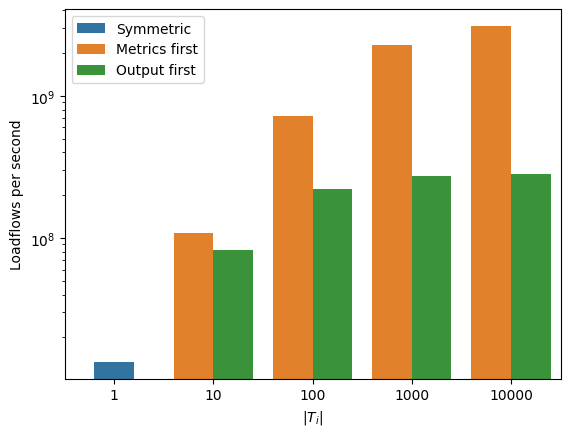}
    \caption{We evaluate two effects, the effect of increasing the size of $T_i$ and computing the topology in metrics-first or output-first mode. These modes are defined in Sec.~\ref{sec:metric_first}. Computed on benchmark grid 2 as presented in Tab.~\ref{table:grids}.}
    \label{fig:injratio}
\end{figure}

In Fig~\ref{fig:injratio}, we compare different numbers of injections per topology, i.e. different sizes of $T_i$. We see a drastic improvement of one to two \ac{OOM} when increasing the size of $T_i$, especially when the injection loop computation is slimmed down in metrics-first mode as discussed in Sec.~\ref{sec:metric_first}. After $|T_i|=1000$ a diminishing return effect starts when the computation time is no longer dominated by the BSDF computation but by the N-1 analysis in the injection loop.

The optimization and preprocessing should be designed to utilize this boost, hence $|T_i|$ should be kept as large as possible.  One trick to assist with this is to replace stub branches (branches that are graph bridges and don't connect the slack, usually connections to power plants) leading out of switched substations by an equivalent injection at the station directly. This reduces the branch connectivity and increases the injection connectivity at the substation. As there are no PV and PQ nodes and no line losses in DC, it is mathematically equivalent to simply move all injections and remove the stub branch.
Having many injections directly in the station furthermore allows to ignore injection combinations that are effectively duplicate. That is, if there are two equally powerful generators, it does not matter which is on which busbar, they are symmetric.

\subsection{Dealing with expensive information extraction functions}
\label{sec:metric_first}
The N-1 matrix in a production grid can easily have $2000^2$ entries or more, which makes it too large to store completely for every topology. Usually, branches which are not overloaded are not of interest. Hence, the N-1 matrix can be stored in a sparsified format, where only the worst cases with respect to their relative load are stored. Using a global top-k operation to filter out the worst k results can be problematic as a single outage case often dominates this statistic, leaving little information from other N-1 cases. Hence, we perform a top-k per N-1 case, concatenate them all and perform a second top-k over this collection.

Top-k is an inefficient and expensive operation on SIMD accelerators \cite{chern2022tpuknnknearestneighbor}, thus step~\ref{aggregate} in the injection module takes a significant amount of time. To deal with this, we implement two running modes of the injection module.

In Fig.~\ref{fig:injratio}, we compare the \textit{output-first} mode to the \textit{metric-first} mode for different $|T_i|$. As presented in Sec.~\ref{sec:gpuapproach}, in output-first mode $agg_i$ is evaluated directly in the injection bruteforce loop versus \textit{metrics-first} mode, where $agg_i$ is evaluated after $t_i^*$ has been found. Note that $|T_i|=1$ is a special case, obviating the need for a bruteforce loop completely; we call this \textit{symmetric} mode. A significant advantage of metrics-first mode can be observed by almost 1 \ac{OOM}, especially with larger $|T_i|$. Slimming down the computations necessary in the injection loop far outweighs the penalty incurred by recomputing the N-0 and N-1 matrices after the loop for the relatively expensive choice of $agg_i$ in our case. 

\subsection{Choice of programming language}
There are multiple ways how to program accelerators: either directly using accelerator specific compilers and languages like CUDA~\cite{cuda} for writing custom kernels, or using abstracted higher-level language that aim to provide compatibility with multiple accelerators at the cost of a bit of performance. We chose the high-level abstraction language jax~\cite{jax2018github} for multiple reasons. First, it offers good performance in benchmarks~\cite{jaxtoast} and works on multiple types of accelerators. The design of the language, especially the functional purity, the similarity to numpy and its vmap transform, improve development ergonomics. Lastly, there are jax based \ac{RL} and \ac{QD} libraries~\cite{qdax, deepmind2020jax} for optimization, which means that the entire optimizer - loadflowsolver interaction can be in jax, enabling the jit compiler optimize the entire loop. We expect the XLA compiler ~\cite{XLA} to produce kernels slightly inferior to hand-written CUDA, however not by a large enough margin to justify the additional developer time.

\subsection{Representing multiple outages}
\label{sec:multi-outages}
Multiple outages are required in two places:  to compute disconnections and to represent outages of three-winding transformers, T-nodes, and busbars in the N-1 computation. These two use-cases differ because for the disconnection computation it is desired to obtain an updated PTDF matrix as all N-1 cases need to be evaluated later on, whereas the N-1 computation itself just needs loadflow results. We investigate the performance of disconnections, as the PTDF update brings a more significant hit to performance than the N-1 multi-outages.

Generally, there are two ways how to compute multiple outages:
\begin{itemize}
    \item Using the LODF formulation in sequence according to Def.~\ref{eq_lodf}.
    \item Using the MODF formulation according to Def.~\ref{eq_modf}.
\end{itemize}

\begin{figure}[tb]
    \includegraphics[width=0.95\linewidth]{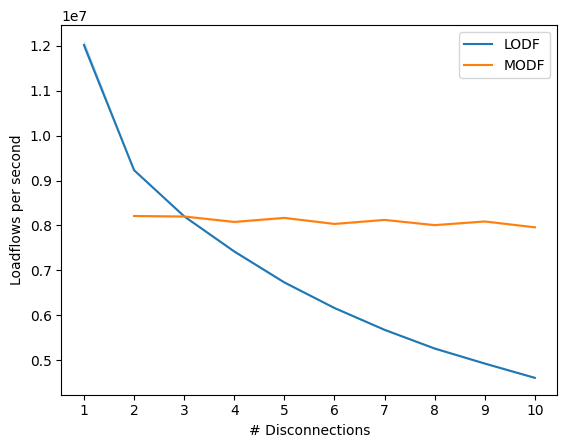}
    \caption{Comparing the repeated LODF and the MODF for different number of disconnections. Computed on benchmark grid 2 with $|T_i|=1$.}
    \label{fig:lodagg_modf}
\end{figure}
We run a benchmark to evaluate the performance difference between the two methods. Figure~\ref{fig:lodagg_modf} shows that for 2 disconnections, the LODF is better than the MODF formulation, while for more disconnections it is preferable to use the MODF. Note that for 1 disconnection, the MODF and LODF are mathematically equivalent and we omit the MODF datapoint. We hypothesize that the comparable poor performance of the MODF method on low number of splits is due to hardware phenomena and poorly optimized code in the inversion of 2x2 matrices. Generally, the stronger scaling properties outweigh the drawbacks, and we opt for MODFs as our implementation for disconnections globally.

The small advantage for 2 disconnections makes an argument to use the repeated LODF in the N-1 computation for 3-winding transformers, as an outage of a 3-winding transformer can be represented as two branch outages and there is a relatively large number of these transformers in our benchmark grids. However, there is a larger drawback. In the setting of injection bruteforcing, it is sufficient to compute the MODF matrices of size ($|t_d| \times |E|$) in the branch module and apply them in the injection module. The repeated LODF formulation requires storing an entire PTDF matrix for every multi-outage, which is of size ($|E| \times |V|$). Hence, the memory footprint of the repeated LODF method for multi-outages in the N-1 computation prevents it from being applied in the branch module, resulting in a drastic slowdown of the injection module. Thus, when $|T_i| > 1$, the MODF is the only viable candidate. The large memory footprint is likely the main reason behind the comparatively poor scaling of the LODF method.

\section{Case Studies}\label{sec:results}

We compare the performance of our solver against multiple baselines and on multiple hardware architectures and grid sizes. For all of our benchmarks, we constrain the number of busbar splits in $t_b$ to three, i.e. it requires three applications of the BSDF (Eq.~\ref{eq_bsdf_application}) to compute the PTDF matrix of the split grid. Furthermore, we set the number of disconnections in $t_d$ to zero because our CPU baseline doesn't support disconnections at the time of writing. Consult Sec.~\ref{sec:multi-outages} for an analysis on the impact of disconnections on the GPU code.

\begin{table}[ht]
\begin{center}
\begin{tabular}{c|c c c c}

    grid & case300 & ACTIVSg2000 & pegase9421 & TSO \\
    & 1 & 2 & 3 & 4 \\
    \hline
    branches & 411 & 3206 & 16049 & 23406 \\
    mon. branches & 411 & 500 & 16049 & 2230 \\
    nodes & 300 & 2000 & 9241 & 14454 \\
    N-1 cases & 385 & 513 & 14384 & 1595 \\
    switchable subs & 15 & 7 & 400 & 64 \\
    with branches & 75 & 37 & 2818 & 483 \\
    with injections & 11 & 3 & 308 & 57 \\
\end{tabular}
\end{center}
\caption{Benchmark grids used.}
\label{table:grids}
\end{table}

We use four grids for benchmarking purposes: 
\begin{enumerate}
    \item A small 300 bus test network \cite{case300}.
    \item A medium size grid based on the ACTIVSg2000 dataset \cite{texasgrid}. We set monitored branches and N-1 cases in this grid to reflect the grid complexity of a use-case at TenneT \cite{viebahn22,viebahn24}.
    \item A large IEEE test case from the pegase project, without any monitored branch or N-1 restrictions. \cite{pegase, josz2016acpowerflowdata}
    \item A production grid used in the operational processes at Elia Group.
\end{enumerate}
For each grid, we define N-1 cases and monitored branches. Furthermore, we designate a subset of substations as switchable substations and display how many branches and injections are connected to these. This represents a topology optimization use case where a single area region is switched, but a larger grid area needs to be monitored for side-effects of the topologies. Furthermore, the larger the set of switchable substations, the lower the effectiveness of the tree-formulation from Sec.~\ref{sec:cpuapproach} as the likelihood of same splits decreases. A summary of the relevant parameters of the test grids is provided in Tab. ~\ref{table:grids}.

We present the following solvers. All experiments using CPU methods were run on 96 AMD EPYC 7763 CPU cores with enabled hyperthreading.
\begin{itemize}
    \item \textbf{pandapower} \cite{pandapower} 2.14.11 using ray \cite{ray} parallelization across topologies and a python for-loop over N-1 cases where \texttt{pp.rundcpp} is called every time. This factorizes a matrix on every N-1 case.
    \item \textbf{pypowsybl} \cite{pypowsybl} 1.9.0 using ray \cite{ray} parallelization across topologies and the native DC security analysis module for N-1 cases. Unfortunately, we could not get it to work for all cases, a fix for the problem is being implemented.
    \item \textbf{Ours, CPU} as presented in Sec.~\ref{sec:cpuapproach}.
    \item \textbf{Ours, GPU} as presented in Sec.~\ref{sec:gpuapproach}.
\end{itemize}

We sample ten topologies and the unsplit configuration for each grid and validate our approaches against the pandapower results up to 1e-6 accuracy. Validating exhaustively was not possible due to the extreme speed difference.

\begin{table}[h]
\begin{center}
\begin{tabular}{c|c c c c}
    grid & 1 & 2 & 3 & 4 \\
    \hline
    pandapower, 96 CPU & 3.0e3 & 1.91e3 & 1.07e2 & 2.0e3 \\
    powsybl, 96 CPU & 7.89e3 & 2.05e3 & err & err \\
    Ours, 96 CPU & 3.0e6 & 1.1e6 & 1.36e6 & 7.9e5 \\
    Ours, 1 H100 & 4.13e8 & 6.14e8 & 1.38e7 & 4.82e7 \\
    Ours, 8 H100 & 2.58e9 & 3.79e9 & 9.94e7 & 3.71e8
\end{tabular}
\end{center}
\caption{Benchmark results of different solvers on the benchmark grids, all results show loadflows per second in scientific notation. $|T_i|$ is set to 100 and topologies with up to 3 splits are evaluated. Entries marked with \textit{err} failed due to a java exception.}
\label{table:benchmark}
\end{table}

% 50Hz
% 1000000 * 1904 / 5.119577169418335 = 371905713.4978834
% 371905713.4978834 * 60 * 60 / 98.32 = 13617377630.109644
% 1000000 * 1904 / 39.52576422691345 = 48171111.60885661
% 48171111.60885661 * 60 * 60 / 6.980 = 24844699397.118023

% case300: 378
% 8 H100 10000000 * 378 / 1.4653027057647705 = 2579671753.234867
% 2579671753.234867 * 60 * 60 / 98.32 = 94455027579.7958
% 1 H100 10000000 * 378 / 9.149068593978882 = 413156810.57279056
% 413156810.57279056 * 60 * 60 / 6.980 = 213089472501.7258

% texas: 498
% 10000000 * 498 / 1.3147318363189697 = 3787844686.216142
% 3787844686.216142 * 60 * 60 / 98.32 = 138692441724.75705
% 10000000 * 498 / 8.112537384033203 = 613864659.6318252
% 613864659.6318252 * 60 * 60 /  6.980 = 316606414709.82385

% pegase: 14384
% 6400 * 14384 / 0.9265925884246826 = 99350675.96051988
% 99350675.96051988 * 60 * 60 / 98.32 =  3637738338.668344
% 6400 * 14384 / 6.692483901977539 = 13755371.151927346
%  13755371.151927346 * 60 * 60 / 6.980 = 7094460766.037026

\begin{table}[h]
\begin{center}
\begin{tabular}{c|c c c c}
    grid & 1 & 2 & 3 & 4 \\
    \hline
    pandapower, 96 CPU & 2.62e6 & 1.67e6 & 9.33e4 & 1.74e6 \\
    powsybl, 96 CPU & 6.95e6 & 1.79e6 & err & err \\
    Ours, 96 CPU & 2.62e9 & 9.59e8 & 1.19e8 & 6.89e8 \\
    Ours, 1 H100 & 2.13e11 & 3.16e11 & 7.09e9 & 2.48e10 \\
    Ours, 8 H100 & 9.45e10 & 1.38e11 & 3.64e9 & 1.36e10
\end{tabular}
\end{center}
\caption{Loadflows per dollar assuming the following costs per machine hour: 96 CPUs: \$4.128, 1 H100: \$6.980, 8 H100: \$98.320. Prices are based on Azure pricings in Q4/2024. Entries marked with \textit{err} failed due to a java exception.}
\label{table:costs}
\end{table}

In Table~\ref{table:benchmark}, the benchmark results are presented in terms of mean loadflows per second averaged across five benchmark runs. Each benchmark ran for ca 15 minutes, assuming that near endless topologies need to be evaluated in an optimization setting. $|T_i|$ was set to 100 to present a conservative scenario and topologies with up to three open couplers, i.e. three BSDF applications, were computed. All benchmarks were run on 64 bit precision on the grids as presented in Tab.~\ref{table:grids}. To facilitate a CPU to GPU comparison, Table~\ref{table:costs} presents the same numbers as Tab.~\ref{table:benchmark} divided by the approximate cost of the machines, based on real Azure US-East on-demand instance pricing.

While powsybl and pandapower reach similar speeds, our CPU method outperforms both by ca 3 \ac{OOM} and the GPU method by ca 5 \ac{OOM}. Pandapower needs to invert a matrix for every N-1 case and is hence expected to be slower. Powsybl in v1.9.0 also uses low-rank updates and should be significantly faster than pandapower, however it is only slightly faster on the grids it can run. According to powsybl developers, a bug in the postprocessing causes extensive memory usage, a later fix might change the numbers drastically. The GPU version is consistently 2 OOMs faster than the CPU version on most grids except for the pegase grid. We assume our GPU code is memory bottlenecked, which is exacerbated by the large grid size. The 8 GPU version outperforms the 1 GPU version by factor 6.2 to 7.7 in all benchmarks, showing near linear scaling. This is expected as there is no cross-device communication during the solving process, mainly the device to host transfer gets more expensive as more results need to be sent back via the device-to-host PCI bus. On grids 2 and 4, where the monitored branches are comparatively few compared to the total number of branches, the PTDF reductions as described in Def.~\ref{eq:ptdf_def} have a strong impact. Looking at loadflows per dollar, the GPU approach consistently outperforms all CPU approaches by at least 1 \ac{OOM}. However, the 8 GPU machine is slightly worse than the 1 GPU machine and is thus only preferable if runtime limits do not permit running on the cheaper machine.

\section{Discussion}\label{sec:Discussion}

We show the effectiveness of using GPUs to accelerate low-rank update based DC loadflow calculations, achieving several orders of magnitudes speed improvement over existing DC solvers in the frame of topology optimization, reaching in the billions of loadflows per second. Using this strategy, gradient-free optimization approaches for topology optimization can be sped up massively, potentially enabling a full-scale topology optimization in close-to-realtime or day-ahead settings. Even exhaustive searches become feasible on smaller grids \cite{viebahn22, viebahn24}. We remark that the DC approximation gives errors up to 10\% against full AC loadflows, trading speed for accuracy, and does not yield voltage magnitudes. Hence, a mixture of AC and DC computations needs to be struck to include all constraints and obtain a full-precision solution.

The solver is also usable in other areas outside of topology optimization where large amounts of DC loadflows need to be computed, such as grid planning applications. Batch sizes in the tens-of-thousand loadflows are necessary to offset the constant overhead of materializing the PTDF matrix and transporting all data to GPU, CPU DC solvers still are preferred if this is not given. Future work could implement Line Closing Distribution Factors, Bus Merge Distribution Factors or the formulation for multiple bus splits/merges from \cite{vandijk2024unifiedalgebraicdeviationdistribution}.

\bibliographystyle{IEEEtran}
\bibliography{main}

\vfill

\end{document}